 \pdfoutput=1
\documentclass[prd,aps,
preprint,superscriptaddress,preprintnumbers,nofootinbib,english]{revtex4-1}
\usepackage[T1]{fontenc}
\usepackage[latin9]{inputenc}
\usepackage{amssymb}
\usepackage{graphicx}

\makeatletter

\providecommand{\tabularnewline}{\\}

\makeatother

\usepackage{babel}
\begin{document}

%\begin{frontmatter}

\title{Baryogenesis from Symmetry Principle}

\author{Chee Sheng Fong}
%\email{fong@if.usp.br}
%\affiliation
\address{Instituto de F\'{\i}sica, 
Universidade de S\~ao Paulo, C.\ P.\ 66.318, 05315-970 S\~ao Paulo, Brazil}
%\ead{fong@if.usp.br}
%\maketitle
\begin{abstract}
In this work, a formalism based on symmetry which allows one to express asymmetries of all the
particles in terms of conserved charges is developed.
The manifestation of symmetry allows one to easily 
determine the viability of a baryogenesis scenario 
and also to identify the different roles played by the symmetry. 
This formalism is then applied to the standard model and its supersymmetric
extension, which constitute two important foundations for constructing models of baryogenesis.
\end{abstract}

\maketitle

%\begin{keyword}
%Baryogenesis, Supersymmetry, Symmetry Principle 
%\end{keyword}

%\end{frontmatter}

\section{Introduction}

The evidences that we live in a matter-dominated Universe are very well-established
\cite{Kolb:1990vq}. While the amount of antimatter is negligible
today, the amount of matter (i.e. baryon) of the Universe has been
determined with great precision by two independent methods. From the
measurement of deuterium abundance originated from Big Bang Nucleosynthesis
(BBN) when the Universe was about a second old (with temperature $T_{{\rm BBN}}\sim{\rm MeV}$),
Ref.~\cite{Cooke:2013cba} quotes the baryon density normalized to
entropic density as $10^{11}Y_{B}^{{\rm BBN}}=8.57\pm0.18$. From
the measurement of temperature anisotropy in the cosmic microwave
background radiation imprinted by acoustic oscillation of photon-baryon
plasma when the Universe was about 380000 years old ($T_{{\rm CMB}}\sim0.3\,{\rm eV}$),
Planck satellite gives $10^{11}Y_{B}^{{\rm CMB}}=8.66\pm0.06$ \cite{Ade:2015xua}.
The impressive agreement between the two measurements 
is a striking confirmation of the standard cosmological model.

In order to account for the cosmic baryon asymmetry, baryogenesis
must be at work before the onset of BBN. 
Although the Standard Model (SM) of particle physics (and cosmology) 
contains all the three ingredients for baryogenesis: 
baryon number violation, $C$ and $CP$ violation,
and the out-of-equilibrium condition \cite{Sakharov:1967dj},
it eventually fails and new physics is called for \cite{Fong:2013wr}. 
Clearly these ingredients are necessary but not sufficient.
Moreover, the early Universe is filled with particles of different
types that interact with each other at various rates, rendering it 
a daunting task to analyze them. In this work, I would
like to advocate the use of \emph{symmetry} as an organizing principle to
analyze such a system.
In particular, I will show that by identifying the symmetries of a
system, one can relate the asymmetries of all the particles to the
corresponding conserved charges without having to take into account 
details of how those particles interact.\footnote{It should be stressed immediately 
that the symmetries do not have to be exact. If a symmetry is approximate, 
the corresponding charge will be quasi-conserved with its evolution described 
by nonequilibrium formalism like Boltzmann equation. In other words, the description 
of the system boils down to identifying \emph{only} the interactions related to 
approximate symmetries.} This should not come
as a surprise since symmetry dictates physics: when we specify a symmetry
and how particles transform under it, the interactions are automatically
fixed. I will first review the formalism in Sec. \ref{sec:Formalism}.
Then the roles of $U(1)$ symmetries are clarified in Sec. \ref{sec:Roles-U(1)}.
In Sec. \ref{sec:The-Standard-Model} and \ref{sec:MSSM} respectively,
I will apply this formalism to the SM and its supersymmetric extension
as they form important bases for constructing models of baryogenesis.
Finally I conclude in Sec. \ref{sec:Conclusions}.

\section{Formalism\label{sec:Formalism}}

Here I will review the formalism that we will use in this work.%
\footnote{The formalism was first introduced by Ref.~\cite{Antaramian:1993nt}
to prove that the generation of hypercharge
asymmetry in a preserved sector implies nonzero baryon asymmetry. 
See also the relevant discussion in Chapter 3.3 of Ref.~\cite{Weinberg:2008zzc}.%
} For a system with $s$ number of symmetries labeled $U(1)_{x}$ and
consisting of $r\geq s$ distinct types of complex particles labeled
$i$ (i.e. not self-conjugate like real scalar or Majorana
fermion) with corresponding chemical potentials $\mu_{i}$ and charges
$q_{i}^{x}$ under $U(1)_{x}$, the most general solution is given
by

\begin{eqnarray}
\mu_{i} & = & \sum_{x}C_{x}q_{i}^{x},\label{eq:chem_pot}
\end{eqnarray}
where $C_{x}$ is some real constant corresponding to $U(1)_{x}$. It
is apparent that Eq.~(\ref{eq:chem_pot}) is the solution for chemical
equilibrium conditions for any possible in-equilibrium interactions
since by definition, the interactions necessarily preserve the symmetry.
Note that symmetry discussed in this work always refers to $U(1)$ 
which characterizes the charge asymmetry between particles and antiparticles. 
The $U(1)_{x}$ can be exact (like gauge symmetry) or approximate (due to small couplings,
and/or suppression by mass scale and/or temperature effects). The diagonal
generators of a nonabelian group do not contribute as long as the
group is not broken \cite{Antaramian:1993nt}. For instance one does
not need to consider conservation of third component of weak isospin
$T_{3}$ before electroweak (EW) phase transition.

Now for each $U(1)_{x}$, according to Noether's theorem there is a conserved current 
and the corresponding conserved charge density can be constructed as

\begin{eqnarray}
n_{\Delta x} & = & \sum_{i}q_{i}^{x}n_{\Delta i},\label{eq:charge_density}
\end{eqnarray}
where $n_{\Delta i}$ is the number density asymmetry for particle $i$.
To proceed we need two further assumptions. Firstly, particle $i$ is assumed 
to participate in fast \emph{elastic} scatterings such that its phase space distribution 
is either Fermi-Dirac $[\exp(E_i-\mu_i)/T + 1]^{-1}$ 
or Bose-Einstein $[\exp(E_i-\mu_i)/T - 1]^{-1}$
for fermion or boson respectively. Secondly, there are 
fast \emph{inelastic} scatterings for particle $i$ and its antiparticle $\bar i$ 
to gauge bosons (which have zero chemical potential) such that $\mu_{\bar i} = - \mu_{i}$.
These two assumptions are justified for instance when the particles have gauge interactions.
Now Eq.~(\ref{eq:charge_density}) can be related to its chemical potential for $\mu_i \ll T$ as follows%
\footnote{The expansion in $\mu_i/T \ll 1$ is justified as long as the number 
asymmetry density is much smaller than its equilibrium number density. For instance with 
$n_{\Delta i}$ the order of the observed baryon asymmetry, the expansion holds when 
the corresponding particle mass over temperature $m_{i}/T \lesssim 20$.}

\begin{eqnarray}
n_{\Delta i} & = & n_{i}-n_{\bar{i}}=\frac{T^{2}}{6}g_{i}\zeta_{i}\mu_{i}.\label{eq:asymmetry_density}
\end{eqnarray}
In the above $g_{i}$ specifies the number of gauge degrees of freedom and 

\begin{eqnarray}
\zeta_{i} & \equiv & \frac{6}{\pi^2}\int_{z_i}^{\infty} dx \,x\sqrt{x^2-z_i^2}
\frac{e^x}{\left(e^x\pm 1\right)^2},\label{eq:zeta}
\end{eqnarray}
with $z_i \equiv m_i/T$. In the relativistic limit $(T\gg m_{i})$, we have $\zeta_{i}=1(2)$
for $i$ a fermion (boson) while in the nonrelativistic limit $(T\ll m_{i})$, we obtain
$\zeta_{i}=\frac{6}{\pi^2} z_i^2 {\cal K}_2(z_i)$ with ${\cal K}_2(x)$ 
the modified Bessel function of type two of order two.
Using Eqs.~(\ref{eq:chem_pot}) and (\ref{eq:asymmetry_density}),
Eq.~(\ref{eq:charge_density}) can be written as

\begin{eqnarray}
n_{\Delta x} & = & \frac{T^{2}}{6}\sum_{y}J_{xy}C_{y},\label{eq:charge_density_C}
\end{eqnarray}
where we have defined the symmetric matrix $J$ as follows

\begin{eqnarray}
J_{xy} & \equiv & \sum_{i}g_{i}\zeta_{i}q_{i}^{x}q_{i}^{y}.\label{eq:J_matrix}
\end{eqnarray}
We can invert Eq.~(\ref{eq:charge_density_C}) to solve for $C_{y}$
in terms of $n_{\Delta x}$ and substituting it into Eq.~(\ref{eq:chem_pot})
and then making use of Eq.~(\ref{eq:asymmetry_density}), we obtain%
\footnote{As long as $r\geq s$ and there are no redundant symmetries, in the
sense that all the symmetries are linearly independent and there is
no rotation in the $s$-dimensional symmetry space that can make all
the $r$ distinct particles uncharged under some $U(1)$, $J$ always
has an inverse. %
}

\begin{eqnarray}
n_{\Delta i} & = & g_{i}\zeta_{i}\sum_{y,x}q_{i}^{y}\left(J^{-1}\right)_{yx}n_{\Delta x}.\label{eq:result}
\end{eqnarray}
Eventually one would like to relate
this to baryon asymmetry i.e. the baryon charge density. 
By substituting Eq.~(\ref{eq:result}) into Eq.~(\ref{eq:charge_density})
for baryon charge density, we have

\begin{eqnarray}
n_{\Delta B} & = & \sum_{y,x}J_{By}\left(J^{-1}\right)_{yx}n_{\Delta x}.\label{eq:baryon_asymmetry}
\end{eqnarray}
Eqs.~(\ref{eq:result}) and (\ref{eq:baryon_asymmetry}) make the
symmetries of the system manifest: the solutions are expressed in term
of conserved charges $n_{\Delta x}$, one for each $U(1)_{x}$ symmetry. 
In fact $\{n_{\Delta x}\}$ forms the \emph{appropriate} basis to describe the system.
While $q_{i}^{x}$ comprises the charges of particle $i$ under $U(1)_{x}$,
$J$ matrix embodies full information of the system 
(all possible interactions consistent with the symmetry are
implicitly taken into account). Notice that calculating
$J$ is particularly \emph{simple} and circumventing the traditional approach of
having to count the number of chemical potentials and determine 
the chemical equilibrium conditions. It is now apparent
that baryogenesis fails ($n_{\Delta B} = 0$) if: 
(I) the system does not possess any symmetry
in which case $C_{x}=0$ for all $x$ in Eq.~(\ref{eq:chem_pot}) 
or; (II) the system possesses only $U(1)_{x}$'s 
which always remain exact such that none develops an
asymmetry in which case $n_{\Delta x}=0$ for all $x$.

For instance, the baryogenesis scenario proposed in Ref.~\cite{Fong:2013gaa}
fails due to the following reasons. In that work, there 
are initially four effective symmetries:
$U(1)_{B/3-L_{\alpha}}(\alpha=\left\{ 1,2,3\right\} )$ and $U(1)_{\tilde{\psi}}$.
During baryogenesis, $U(1)_{B/3-L_{\alpha}}$ is always conserved 
i.e. $n_{\Delta(B/3-L_{\alpha})}=0$
while a large enough $CP$ asymmetry at the TeV scale 
requires fast $U(1)_{\tilde{\psi}}$ violation i.e. $C_{\tilde{\psi}}=0$.
As a result, $n_{\Delta B}=0$.

\section{The roles of $U(1)$ symmetries\label{sec:Roles-U(1)}}

In general, the reaction rate of a process $\gamma$
in the early Universe is temperature-dependent $\Gamma_{\gamma}(T)$.
At each range of temperature $T^{*}$, by comparing $\Gamma_{\gamma}(T^*)$
to the expansion rate of the Universe $H(T^*)$, we can categorize the
reactions into three types \cite{Fong:2010qh,Fong:2010bv}: 
(i) $\Gamma_{\gamma}\left(T^{*}\right)\gg H\left(T^{*}\right)$;
(ii) $\Gamma_{\gamma}\left(T^{*}\right)\ll H$; 
(iii) $\Gamma_{\gamma}\left(T^{*}\right)\sim H\left(T^{*}\right)$.
The reactions of type (i) are fast enough to establish chemical equilibrium
and are implicitly `resummed' in the $J$ matrix in Eq.~(\ref{eq:J_matrix}).
The reactions of type (ii) either do not occur or proceed slow enough.
The former is due to exact symmetry like gauge symmetry 
while the latter is due to small couplings, and/or suppression by mass scale
and/or temperature effects. 
Finally the reactions of type (iii) should be described by nonequilibrium formalism like
Boltzmann equation in order to obtain quantitative prediction. In
this work, the effective symmetries concern both reactions of types
(ii) and (iii). In particular gauge symmetry always belongs to type (ii) and 
can play an interesting role as `\emph{messenger}'. If an approximate
symmetry belongs to type (ii), it can acquire a role as a `\emph{messenger}'
or `\emph{preserver}' while if it is of type (iii), it can act
as `\emph{creator/destroyer}'. 

To understand the roles of $U(1)$ alluded to above, it
is illuminating to group the charges as follows. Among
all the charges $U=\left\{ n_{\Delta x}\right\} $, there is a subset
$U_{0}=\left\{ n_{\Delta a}\right\} $ where the net charges vanish $n_{\Delta a}=0$.
In this case, we can remove them from the beginning and left with
$\tilde{U}=U-U_{0}=\left\{ n_{\Delta m}\right\} $ to describe the
system. From Eq.~(\ref{eq:charge_density_C}), 
we have a set of linear equations $n_{\Delta a}=\sum_{b}J_{ab}C_{b}+\sum_{m}J_{am}C_{m}=0$,
which allows us to solve for $C_{a}$ in terms of $C_{m}$.% 
\footnote{We use $a,b,...$ to label the charges in $U_{0}$
and $m,n,...$ to label the charges in $\tilde{U}$.%
} 
After eliminating $C_a$, the number density asymmetry for 
particle $i$ can be expressed as

\begin{eqnarray}
n_{\Delta i} & = & g_{i}\zeta_{i}\sum_{m,n} \tilde{q}_{i}^{m}
\left(\tilde{J}^{-1}\right)_{mn}n_{\Delta n}.\label{eq:result2}
\end{eqnarray}
where we have defined 
$\tilde{q}_{i}^{m} \equiv q_{i}^{m}-\sum_{a,b}q_{i}^{a}\left(J^{-1}\right)_{ab}J_{bm}$
and
$\tilde{J}_{mn} \equiv J_{mn}-\sum_{a,b}J_{ma}\left(J^{-1}\right)_{ab}J_{bn}.$ 
Substituting Eq.~(\ref{eq:result2}) into Eq.~(\ref{eq:charge_density})
for baryon charge density, we get

\begin{eqnarray}
n_{\Delta B} & = & \sum_{m,n}\left[J_{Bm}-\sum_{a,b}J_{Ba}\left(J^{-1}\right)_{ab}J_{bm}\right]\left(\tilde{J}^{-1}\right)_{mn}n_{\Delta n}.\label{eq:baryon_asymmetry2}
\end{eqnarray}
The equation above can be succinctly written as 
$n_{\Delta B} = \sum_{m,n}\tilde{J}_{Bm} \left(\tilde{J}^{-1}\right)_{mn}n_{\Delta n}$ 
but it is elucidating to keep it as it is:
the two terms in the square bracket of Eq.~(\ref{eq:baryon_asymmetry2}) represent two different
types of contributions to the baryon asymmetry. The first term is
the direct contribution of $\tilde{U}$ sector to the baryon asymmetry
while the second term is the contribution of $\tilde{U}$ sector through
$U_{0}$ (the messenger sector). Hence even if $\tilde{U}$ sector
does not carry baryon charge $J_{Bm}=0$, as long as it carries
charges in the messenger sector $J_{bm}\neq0$, and some baryons 
also carry charges in the messenger sector $J_{Ba}\neq0$, 
we will have $n_{\Delta B}\neq0$.
Here $\tilde{U}$ sector can play two roles: as creator/destroyer
or preserver of asymmetries depending on their rates as discussed
in the beginning of this section. In short, the roles of $U(1)$
symmetries in baryogenesis can be concisely stated as follows:
\begin{enumerate}
\item \emph{Creator/destroyer}: type (iii) reaction with an approximate $U(1)_{m}$.
The dynamical violation of $U(1)_{m}$ results in the development
of $n_{\Delta m}\neq0$ from $n_{\Delta m}=0$.
As mentioned earlier, quantitative prediction requires one to solve dynamical equation like
Boltzmann equation for $n_{\Delta m}$ and the generated asymmetry depends 
on the rates of \emph{creation} and \emph{washout}. 
\item \emph{Preserver}: type (ii) reaction with $U(1)_{m}$ and $n_{\Delta m}\neq0$. 
The symmetry prevents the asymmetry from being washed out.
The lightest electrically neutral particle in this sector can 
be a good (asymmetric) dark matter candidate.
\item \emph{Messenger}: type (ii) reaction with $U(1)_{a}$ and $n_{\Delta a}=0$. 
Further requirement is that at least some particles in $U(1)_{m}$ 
(of the preserver or the creator/destroyer) 
and some baryons need to be charged under $U(1)_{a}$ such that a nonzero asymmetry
in $U(1)_{m}$ induces a nonzero baryon asymmetry through $U(1)_{a}$ conservation.
\end{enumerate}

In the SM, conservations of hypercharge $U(1)_{Y}$ and electric charge $U(1)_{Q}$ 
ensure $n_{\Delta Y} = n_{\Delta Q} =0$. Hence they play the role 
of messenger respectively before and after EW phase transition (EWPT) at $T_{\rm EWPT}$. 
Eq.~(\ref{eq:baryon_asymmetry2}) is the generalization of the result
of Ref.~\cite{Antaramian:1993nt} which shows that a preserved sector
which carries nonzero hypercharge asymmetry implies nonzero baryon
asymmetry (set $a=b=Y$ in the second term in Eq.~(\ref{eq:baryon_asymmetry2})).
We can readily extend this result to post-EW-sphaleron baryogenesis
scenario \cite{Babu:2006xc} where $U(1)_{Q}$ plays the role of messenger.
In this case, baryon asymmetry cannot be erased by fast $B$-violating
interactions as long as there is a preserved sector carrying nonzero
electric charge asymmetry. Of course, phenomenological constraint
will require that the electric charge asymmetry to decay away before BBN.

\section{The Standard Model\label{sec:The-Standard-Model}}

\begin{figure}
\includegraphics[scale=0.6]{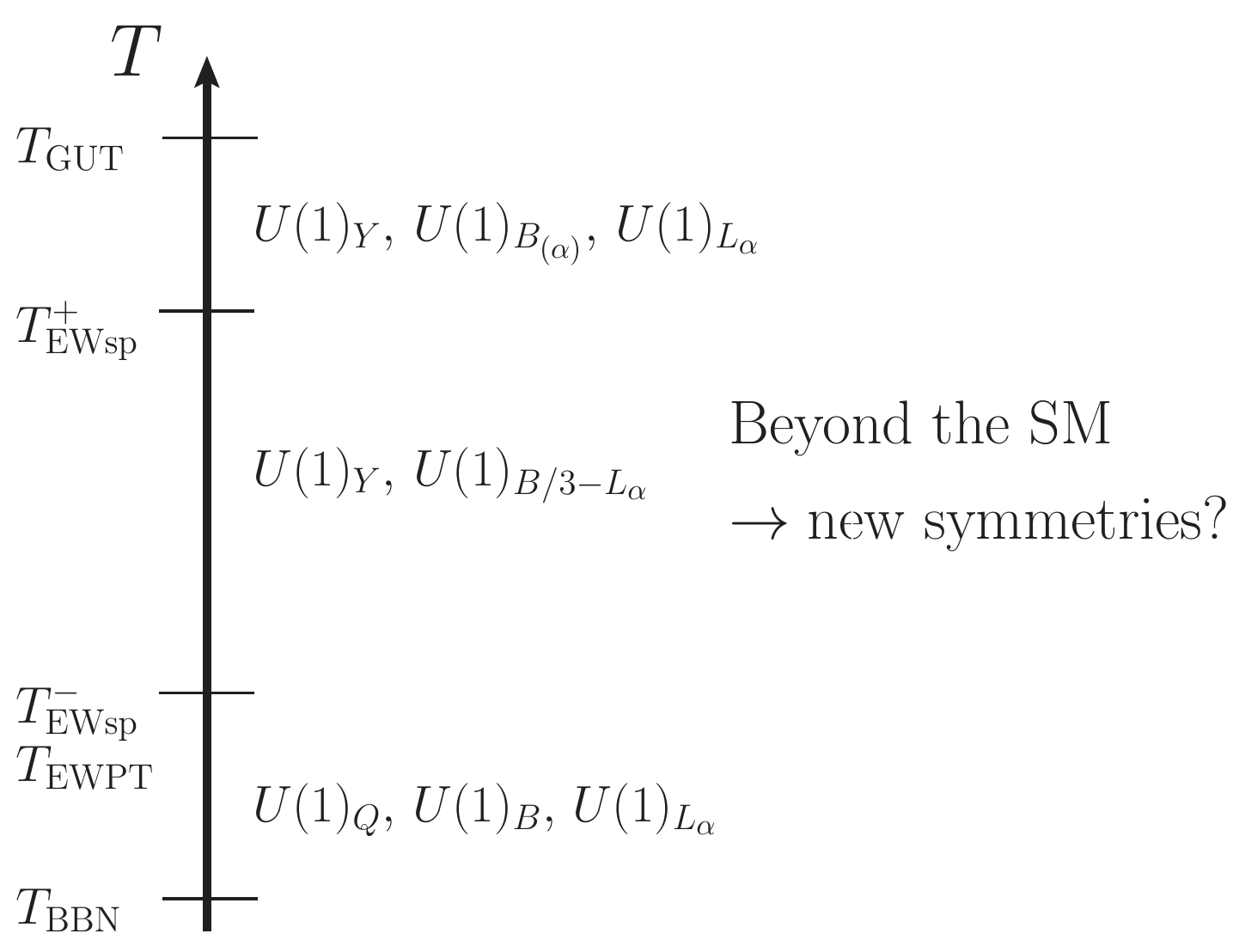}
\caption{\label{fig1}Symmetries of the SM \label{fig:SM_symmetries} in the early Universe
in between BBN and grand unified theory scale ($T_{\rm GUT} \sim 10^{16}$ GeV). 
$T_{{\rm EWsp}}^{-}<T<T_{{\rm EWsp}}^{+}$
is the range of temperature where EW sphalerons are in thermal equilibrium.
At very high temperature $T\gtrsim T_{{\rm EWsp}}^{+}$,
some of the interactions due to quark intergeneration mixing can become
ineffective, resulting in baryon flavor conservation $U(1)_{B_{\alpha}}$.
While $U(1)_{Y}$ and $U(1)_{Q}$ are gauge symmetries which have to be exact;
$U(1)_{B_{\left(\alpha\right)}}$,
$U(1)_{L_{\alpha}}$ and $U(1)_{B/3-L_{\alpha}}$ are global symmetries
which can be broken dynamically, providing the avenue for baryogenesis.
}

\end{figure}
First let us define the $U(1)_{x}-SU(N)-SU(N)$ mixed anomaly (coefficient) as 
$A_{xNN}\equiv\sum_{i}c_{2}\left(R\right)g_{i}q_{i}^{x}$
where $c_{2}\left(R\right)$ is the quadratic Casimir operator in
representation $R$ of $SU(N)$ with $c_{2}\left(R\right)=\frac{1}{2}$
in the fundamental representation and $c_{2}\left(R\right)=N$ in
the adjoint representation. Here $g_{i}$ is the degeneracy of particle
$i$ of charge $q_{i}^{x}$ in a given representation. In the following,
for $N=2$, we always refer to weak $SU(2)_{L}$ while
for $N=3$, color $SU(3)_{c}$. 

The SM Yukawa sector is described by

\begin{eqnarray}
-{\cal L}_{Y} & = & \left(y_{u}\right)_{\alpha\beta}\overline{Q_{\alpha}}\epsilon H^{*}U_{\beta}+\left(y_{d}\right)_{\alpha\beta}\overline{Q_{\alpha}}HD_{\beta}+\left(y_{e}\right)_{\alpha\beta}\overline{\ell_{\alpha}}HE_{\beta}+{\rm H.c.},\label{eq:SM_Lag}
\end{eqnarray}
where $\alpha,\beta=\left\{ 1,2,3\right\} $ are fermion family indices
and the $SU(2)_{L}$ contraction is shown explicitly with antisymmetric
tensor $\epsilon_{01}=-\epsilon_{10}=1$ and $\epsilon_{ii}=0$ while
the $SU(3)_{c}$ contraction is left implicit.
In Eq.~(\ref{eq:SM_Lag}), $Q_{\alpha}$, $\ell_{\alpha}$, $H$ are respectively
the left-handed quark, left-handed lepton and Higgs $SU(2)_{L}$ doublets
while the right-handed quark and lepton singlets are $U_{1}=u$, $D_{1}=d$,
$E_{1}=e$ and so on. Besides $U(1)_{Y}$ or $U(1)_{Q}$,
it is well-known that there are baryon $U(1)_{B}$ and lepton flavors
$U(1)_{L_{\alpha}}$ as accidental symmetries. The relevant $U(1)$ charges 
are listed in Table \ref{tab:SM_charges_before_EWPT}. $U(1)_{Y}$ and $U(1)_{Q}$
as gauge symmetries are ensured to be anomaly-free. On the other hand,
$U(1)_{B}$ and $U(1)_{L_{\alpha}}$ both have $SU(2)_{L}$ mixed anomaly
with $A_{B22}=A_{L_{\alpha}22}=\frac{N_{f}}{2}$ where $N_{f}$ is
the number of fermion family ($N_f = 3$ in the SM). As a
result of the anomaly, $B$ and $L_{\alpha}$ are both violated by sphaleron-mediated
dimension-6 operator ${\cal O}_{{\rm EWsp}}=\sum_{\alpha}\left(QQQ\ell\right)_{\alpha}$ 
(here onwards, these interactions
will be referred to as EW sphalerons) \cite{'tHooft:1976up}. Nonetheless, 
one can form an anomaly-free charge combination $U(1)_{\left(B-L\right)_{\alpha}}$
respected by ${\cal O}_{{\rm EWsp}}$. Although exponentially suppressed
today \cite{'tHooft:1976up}, the EW sphalerons are in thermal equilibrium
in the temperature range $T_{{\rm EWsp}}^{-}<T<T_{{\rm EWsp}}^{+}$
with $T_{{\rm EWsp}}^{-}\sim100$ GeV and $T_{{\rm EWsp}}^{+}\sim10^{12}$
GeV \cite{Kuzmin:1985mm,Bento:2003jv}. For most of the epoch in the early
Universe, quark intergeneration mixing violates baryon flavors and
hence the exact symmetries are instead $U(1)_{\Delta_{\alpha}}$ 
with $\Delta_{\alpha}\equiv B/3-L_{\alpha}$.%
\footnote{In the SM, in the absence of right-handed neutrinos, lepton flavors
are conserved. In beyond the SM scenario, fast lepton flavor violation could
occur see for e.g. \cite{AristizabalSierra:2009mq}.%
} Outside this temperature window, $B$ and $L_{\alpha}$ are effectively conserved. 
In the SM before EWPT, $U(1)_{\Delta_{\alpha}}$
can act as both creator and preserver while $U(1)_{Y}$ is a
messenger. After EWPT, the role of $U(1)_{\Delta_{\alpha}}$ is taken
over by $U(1)_{B}$ while the messenger becomes $U(1)_{Q}$. 
More often than not, when one considers scenario beyond the SM, there are
new symmetries (see Sec. \ref{sec:MSSM}) which can play the roles
of $U(1)$ discussed before. The symmetries of the SM in the early Universe 
is summarized in Fig.~\ref{fig1}.

\begin{table}
\begin{tabular}{|c|c|c|c|c|c|c|c|}
\hline 
$i=$ & $Q_{\alpha}$ & $U_{\alpha}$ & $D_{\alpha}$ & $\ell_{\alpha}$ & $E_{\alpha}$ & $H$ & $H'$\tabularnewline
\hline 
\hline 
$q_{i}^{\Delta_{\alpha}}$ & $\frac{1}{9}$ & $\frac{1}{9}$ & $\frac{1}{9}$ & $-1$ & $-1$ & $0$ & $0$\tabularnewline
\hline 
$q_{i}^{Y}$ & $\frac{1}{6}$ & $\frac{2}{3}$ & $-\frac{1}{3}$ & $-\frac{1}{2}$ & $-1$ & $\frac{1}{2}$ & $\frac{1}{2}$\tabularnewline
\hline 
\hline 
$q_{i}^{B}$ & $\frac{1}{3}$ & $\frac{1}{3}$ & $\frac{1}{3}$ & $0$ & $0$ & $0$ & $0$\tabularnewline
\hline 
$q_{i}^{L_{\alpha}}$ & $0$ & $0$ & $0$ & $1$ & $1$ & $0$ & $0$\tabularnewline
\hline 
$g_{i}$ & $3\times2$ & $3$ & $3$ & $2$ & $1$ & $2$ & $2(N_{H}-1)$\tabularnewline
\hline 
\end{tabular}

\caption{\label{tab:SM_charges_before_EWPT}The list of SM fields, their $U(1)$
charges $q_{i}^{x}$ and gauge degrees of freedom $g_{i}$ with fermion
family index $\alpha$. Here $N_{H}-1$ is number of extra pairs
of Higgses $H'$ with the assumption that they maintain chemical equilibrium
with the SM Higgs $H$. }
\end{table}

\subsection{Some specific cases}

Let us define the vectors $q_{i}^{T}\equiv\left(q_{i}^{\Delta_{\alpha}},q_{i}^{Y}\right)$
and $n^{T}\equiv\left(n_{\Delta_{\alpha}},n_{\Delta Y}\right)$. 
Consider first the temperature regime $T\sim10^{4}$ GeV
when all Yukawa interactions are in thermal equilibrium and 
all particles are relativistic. In this case,
the $J$ matrix is easily determined from Eq.~(\ref{eq:J_matrix})
to be (here we express in its inverse) 

\begin{eqnarray}
J^{-1} & = & \frac{1}{3\left(198+39N_{H}\right)} \nonumber \\
& &\times\left(\begin{array}{cccc}
222+35N_{H} & 4\left(6-N_{H}\right) & 4\left(6-N_{H}\right) & -72\\
4\left(6-N_{H}\right) & 222+35N_{H} & 4\left(6-N_{H}\right) & -72\\
4\left(6-N_{H}\right) & 4\left(6-N_{H}\right) & 222+35N_{H} & -72\\
-72 & -72 & -72 & 117
\end{array}\right),
\end{eqnarray}
where $N_{H}-1$ is number of extra pairs of Higgses $H'$ with the
assumption that they maintain chemical equilibrium with the SM Higgs
$H$. Using the matrix above, $n_{\Delta i}$ can be expressed in terms of conserved
charge densities through Eq.~(\ref{eq:result}). In particular, setting
$N_{H}=1$ and $n_{\Delta Y}=0$, we obtain lepton asymmetries $n_{\Delta\ell_{\alpha}}$
and Higgs asymmetries $n_{\Delta H}$ in terms of $n_{\Delta_{\alpha}}$
which are in agreement with Ref.~\cite{Nardi:2006fx}. We can further
consider cases at higher temperature when $e$ Yukawa interactions
are out-of-equilibrium in which we gain a chiral $U(1)_{e}$. Formally we
can create another conserved charge $n_{\Delta e}$ and determine $J$ which 
is now a $5\times5$ matrix. However if we were to take $n_{\Delta e}=0$, in practice,
we can just set $\zeta_{e}=0$ from the beginning to exclude its contribution. 
Next consider the case when both $u$ and $d$ Yukawa interactions
are out-of-equilibrium. Since both $U(1)_{u}$ and $U(1)_{d}$ 
have $SU(3)_{c}$ mixed anomaly, no effective symmetry is gained. 
Nevertheless $u$ and $d$ are now indistinguishable under QCD sphalerons and we simply
have to set $q_{u}^{Y}=q_{d}^{Y}=\frac{1}{2}\left(\frac{2}{3}-\frac{1}{3}\right)=\frac{1}{6}$.
It is straightforward to consider further cases 
and the results of Refs.~\cite{Nardi:2006fx,Abada:2006ea} are verified.

\subsection{Relation between $B$ and $B-L$}

An important quantity for high scale baryogenesis (occurs at $T>T_{{\rm EWsp}}^{-}$)
is the relation between the $B$ and $B-L$ charge densities during
the time when EW sphalerons freeze out. For simplicity, we will assume 
that all particles are relativistic although particle decoupling effects \cite{Inui:1993wv,Chung:2008gv}
can be straightforwardly taken into account by considering generic form of $\zeta_i$ 
as in Eq.~(\ref{eq:zeta}). Assuming $T_{{\rm EWsp}}^{-}>T_{{\rm EWPT}}$,
the conserved charges are $n_{\Delta Y}$ and $n_{\Delta(B-L)}\equiv\sum_{\alpha}n_{\Delta_{\alpha}}$.
By defining the vectors $q_{i}^{T}\equiv\left(q_{i}^{B-L},q_{i}^{Y}\right)$
and $n^{T}\equiv\left(n_{\Delta(B-L)},n_{\Delta Y}\right)$ and keeping
the number of fermion family as $N_{f}$ with $N_{H}$ pairs of Higgses,
we obtain the following

\begin{eqnarray}
J^{-1} & = & \frac{1}{N_{f}\left(22N_{f}+13N_{H}\right)}\left(\begin{array}{cc}
10N_{f}+3N_{H} & -8N_{f}\\
-8N_{f} & 13N_{f}
\end{array}\right).
\end{eqnarray}
Setting $n_{\Delta Y}=0$, we have from Eq.~(\ref{eq:baryon_asymmetry})

\begin{eqnarray}
n_{\Delta B} & = & \frac{4\left(2N_{f}+N_{H}\right)}{22N_{f}+13N_{H}}n_{\Delta(B-L)}.
\end{eqnarray}
On the other hand, assuming $T_{{\rm EWsp}}^{-}<T_{{\rm EWPT}}$,
we need to consider the components of $SU(2)_{L}$ doublets 
and use $Q$ in place of $Y$ as in Table \ref{tab:SM_charges_after_EWPT}.
Doing so we obtain

\begin{eqnarray}
J^{-1} & = & \frac{1}{2N_{f}\left[24N_{f}+13\left(2+N_{H}\right)\right]}\left(\begin{array}{cc}
2\left(6+8N_{f}+3N_{H}\right) & -8N_{f}\\
-8N_{f} & 13N_{f}
\end{array}\right).
\end{eqnarray}
Now setting $n_{\Delta Q}=0$, we have from Eq.~(\ref{eq:baryon_asymmetry})

\begin{eqnarray}
n_{\Delta B} & = & \frac{4\left(2+2N_{f}+N_{H}\right)}{24N_{f}+13\left(2+N_{H}\right)}n_{\Delta(B-L)}.
\end{eqnarray}
The results above agree with Ref.~\cite{Harvey:1990qw} albeit obtained from 
simpler derivation based on symmetry principle.

\begin{table}
\begin{tabular}{|c|c|c|c|c|c|c|c|c|c|}
\hline 
$i=$ & $U_{\alpha,L}$ & $D_{\alpha,L}$ & $U_{\alpha}$ & $D_{\alpha}$ & $\nu_{\alpha,L}$ & $E_{\alpha,L}$ & $E_{\alpha}$ & $W^{+}$ & $H'^{+}$\tabularnewline
\hline 
\hline 
$q_{i}^{\Delta_{\alpha}}$ & $\frac{1}{9}$ & $\frac{1}{9}$ & $\frac{1}{9}$ & $\frac{1}{9}$ & $-1$ & $-1$ & $-1$ & $0$ & $0$\tabularnewline
\hline 
$q_{i}^{Q}$ & $\frac{2}{3}$ & $-\frac{1}{3}$ & $\frac{2}{3}$ & $-\frac{1}{3}$ & $0$ & $-1$ & $-1$ & $1$ & $1$\tabularnewline
\hline 
\hline 
$q_{i}^{B}$ & $\frac{1}{3}$ & $\frac{1}{3}$ & $\frac{1}{3}$ & $\frac{1}{3}$ & $0$ & $0$ & $0$ & $0$ & $0$\tabularnewline
\hline 
$q_{i}^{L}$ & $0$ & $0$ & $0$ & $0$ & $1$ & $1$ & $1$ & $0$ & $0$\tabularnewline
\hline 
$g_{i}$ & $3$ & $3$ & $3$ & $3$ & $1$ & $1$ & $1$ & $3$ & $N_{H}-1$\tabularnewline
\hline 
\end{tabular}

\caption{\label{tab:SM_charges_after_EWPT}Similar to Table \ref{tab:SM_charges_before_EWPT}
but for field components after EWPT where we use subscript `$L$'
to denote the left-handed fields which participate in weak interaction.}
\end{table}

\section{The Minimal Supersymmetric Standard Model\label{sec:MSSM}}

Here we consider a well-motivated extension to the SM which is the minimal
supersymmetric SM (MSSM). The MSSM superpotential is given by

\begin{eqnarray}
W & = & \mu_{H}H_{u}\epsilon H_{d}
+\left(y_{u}\right)_{\alpha\beta}Q_{\alpha}\epsilon H_{u}U_{\beta}^{c} \nonumber \\
& &+\left(y_{d}\right)_{\alpha\beta}Q_{\alpha}\epsilon H_{d}D_{\beta}^{c}
+\left(y_{e}\right)_{\alpha\beta}\ell_{\alpha}\epsilon H_{d}E_{\beta}^{c},\label{eq:MSSM_superpotential}
\end{eqnarray}
where all the fields above stand for \emph{left-chiral superfields}.
One observes that the superpotential has an $R$ symmetry $U(1)_{R}$ 
for e.g. with $q^{R}\left(H_{d}\right)=q^{R}\left(\ell_{\alpha}\right)=q^{R}\left(U_{\alpha}^{c}\right)=-q^{R}\left(E_{\alpha}^{c}\right)=2$
and the rest of the fields having zero charges.%
\footnote{Note that the $R$-symmetry is preserved also with $R$-parity violating
terms as well as in supersymmetric type-I seesaw with right-handed
neutrino chiral superfields $N_{i}^{c}$ having $R\left(N_{i}^{c}\right)=0$.%
} The $R$ symmetry has mixed anomalies $A_{R33}=3-N_{f}$
and $A_{R22}=2-N_{f}$ where $N_{f}$ is the number of fermion family.
With $N_{f}=3$, there is only $A_{R22}=-1$ anomaly. Thus one can 
form an anomaly-free charge combination as follows

\begin{eqnarray}
\overline{R} & \equiv & R+\frac{2}{3c_{BL}}\left(c_{B}B+c_{L}L\right),\label{eq:Rbar}
\end{eqnarray}
with $c_{BL}\equiv c_{B}+c_{L}$ any number. Notice that $\overline{R}$ is exactly
conserved by Eq.~(\ref{eq:MSSM_superpotential}). Further setting
$\mu_{H}=0$, we gain an anomalous global $U(1)_{PQ}$ for e.g. 
with $-q^{PQ}\left(Q_{\alpha}\right)=q^{PQ}\left(\ell_{\alpha}\right)
=q^{PQ}\left(H_{u}\right)=q^{PQ}\left(H_{d}\right)=1$,
$q^{PQ}\left(E_{\alpha}^{c}\right)=-2$ and the rest of the fields
having zero charges. One can verify that $U(1)_{PQ}$ is anomalous
with $A_{PQ33}=-N_{f}$ and $A_{PQ22}=-N_{f}+N_{H}$. With $N_{f}=3$
and $N_{H}=1$, the $A_{PQ22}$ anomaly-free charge combination is 

\begin{eqnarray}
\overline{P} & \equiv & \frac{3}{4}c_{BL}PQ+c_{B}B+c_{L}L.\label{eq:Pbar}
\end{eqnarray}
In order to cancel the $A_{PQ33}$ anomaly, we need another mixed
$SU(3)_{c}$ anomalous symmetry. For instance, when the $u$ Yukawa
interactions are out-of-equilibrium, we gain an anomalous chiral symmetry
$U(1)_{u^{c}}$ with $A_{u^{c}33}=q^{u^{c}}/2$. The anomaly-free
charge combination is

\begin{eqnarray}
\overline{\chi_{u^{c}}} & \equiv & \overline{P}+\frac{9}{2}c_{BL}u^{c}/q^{u^{c}}.\label{eq:chibar}
\end{eqnarray}
The $U(1)$ charges of the superfields are listed in Table \ref{tab:charges_MSSM}.
The anomalous $U(1)_{R}$ and $U(1)_{PQ}$ discussed above were first
studied in Ref.~\cite{Ibanez:1992aj} and were shown to be effective at 
$T\gtrsim10^{7}$ GeV when the interactions mediated by
weak scale $\mu_{H}$, soft trilinear couplings and gaugino masses
are out-of-equilibrium. While these symmetries impart only order of
one effects in the standard supersymmetric leptogenesis \cite{Fong:2010qh},
it significantly enhances the efficiency of soft leptogenesis \cite{Fong:2010bv}. 

Finally it should be remarked that $c_B$ and $c_L$ can be chosen 
at will depending on the baryogenesis model under consideration.
For instance, considering a model which violates lepton number through
${\cal O}_{L}=(\ell_{\alpha}\epsilon H_{u})^2$, we can choose $c_{B}=-5c_{L}/3$ 
such that $\overline{R}$ and $\overline{P}$ are conserved by ${\cal O}_{L}$.
As another example, considering a model which violates baryon
number through ${\cal O}_{B}=U_{\alpha}^{c}D_{\beta}^{c}D_{\delta}^{c}$,
a good choice is $c_{B}=0$ and $c_{L}\neq0$ such that $\overline{R}$ and
$\overline{P}$ are conserved by ${\cal O}_{B}$. Choosing $c_{B}=c_{L}$, 
the results obtained are in disagreement with Ref.~\cite{Ibanez:1992aj} 
due to sign error of gaugino chemical potential in their Eq.(3.3).%
\footnote{For instance we have $n_{\Delta B}=6N_{f}\left(2n_{\Delta Q}+n_{\Delta\widetilde{G}}\right)$
instead of $n_{\Delta B}=2\left[6N_{f}n_{\Delta Q}-\left(4N_{f}-9\right)n_{\Delta\widetilde{G}}\right]$.
Since the derivation here is solely based on symmetries of the system, this
mistake will not occur.%
} 

\begin{table}
{\footnotesize }%
\begin{tabular}{|c|c|c|c|c|c|c|c|}
\hline 
{ $i=$} & { $Q_{a}$} & { $U_{a}^{c}$} & { $D_{a}^{c}$} & { $\ell_{\alpha}$} & { $E_{\alpha}^{c}$} & { $H_{u}$} & { $H_{d}$}\tabularnewline
\hline 
\hline 
{ $q_{i}^{\Delta_{\alpha}}$} & { $\frac{1}{9}$} & { $-\frac{1}{9}$} & { $-\frac{1}{9}$} & { $-1$} & { $1$} & { $0$} & { $0$}\tabularnewline
\hline 
{ $q_{i}^{Y}$} & { $\frac{1}{6}$} & { $-\frac{2}{3}$} & { $\frac{1}{3}$} & { $-\frac{1}{2}$} & { $1$} & { $\frac{1}{2}$} & { $-\frac{1}{2}$}\tabularnewline
\hline 
{ $q_{i}^{\overline{R}}$} & { $\frac{2c_{B}}{9c_{BL}}$} & { $2-\frac{2c_{B}}{9c_{BL}}$} & { $-\frac{2c_{B}}{9c_{BL}}$} & { $2+\frac{2c_{L}}{3c_{BL}}$} & { $-2-\frac{2c_{L}}{3c_{BL}}$} & { $0$} & { $2$}\tabularnewline
\hline 
\hline 
{ $q_{i}^{\overline{P}}$} & { $\frac{c_{B}}{3}-\frac{3c_{BL}}{4}$} & { $-\frac{c_{B}}{3}$} & { $-\frac{c_{B}}{3}$} & { $c_{L}+\frac{3c_{BL}}{4}$} & { $-c_{L}-\frac{3c_{BL}}{2}$} & { $\frac{3c_{BL}}{4}$} & { $\frac{3c_{BL}}{4}$}\tabularnewline
\hline 
{ $q_{i}^{B}$} & { $\frac{1}{3}$} & { $-\frac{1}{3}$} & { $-\frac{1}{3}$} & { $0$} & { $0$} & { $0$} & { $0$}\tabularnewline
\hline 
{ $q_{i}^{L}$} & { $0$} & { $0$} & { $0$} & { $1$} & { $-1$} & { $0$} & { $0$}\tabularnewline
\hline 
{ $q_{i}^{PQ}$} & { $-1$} & { $0$} & { $0$} & { $1$} & { $-2$} & { $1$} & { $1$}\tabularnewline
\hline 
{ $q_{i}^{R}$} & { $0$} & { $2$} & { $0$} & { $2$} & { $-2$} & { $0$} & { $2$}\tabularnewline
\hline 
{ $g_{i}$} & { $3\times2$} & { $3$} & { $3$} & { $2$} & { $1$} & { $2$} & { $2$}\tabularnewline
\hline 
\end{tabular}{\footnotesize \par}

\caption{\label{tab:charges_MSSM} The $U(1)$ charges of left-handed chiral
superfields. All gauginos $\widetilde{G}$, $\widetilde{W}$
and $\widetilde{B}$ have both $R$ and $\overline{R}$ charges equal
1. Since all fermions in chiral superfields have $R$ charges one
less than that of bosons i.e. $R\left({\rm fermion}\right)=R\left({\rm boson}\right)-1$,
the differences between number density asymmetries of bosons and
fermions are equal to that of gauginos.}
\end{table}

\section{Conclusions\label{sec:Conclusions}}

The use of symmetry principle in analyzing the early Universe system
allows all the particle asymmetries to be
expressed in terms of conserved charges corresponding to the symmetries. 
These charges form the appropriate basis to describe the system.
Besides its simplicity i.e. without having to resort to details of how the particles
interact, this method serves as a powerful tool in accessing the viability of 
a baryogenesis scenario. In addition, the roles of $U(1)$ symmetries as 
creator/destroyer, preserver or messenger become apparent, 
rendering it easier to construct interesting models of baryogenesis.

\section*{Acknowledgements}

%\begin{acknowledgments}
This work is supported by Funda\c{c}\~ao de Amparo \`a Pesquisa do
Estado de S\~ao Paulo (FAPESP) 2013/13689-7. 
%\end{acknowledgments}

\end{document}